\documentstyle[12pt,epsf]{article}
\begin{document}

\title{On the correspondence between the classical and quantum gravity}

\author{~Kirill~A.~Kazakov\thanks{E-mail: $kirill@theor.phys.msu.su$}}

\maketitle

\begin{center}
{\em Moscow State University, Physics Faculty,\\
Department of Theoretical Physics.\\
$117234$, Moscow, Russian Federation}
\end{center}

\begin{abstract}
The relationship between the classical and quantum theories of gravity
is reexamined. The value of the gravitational potential defined with the
help of the two-particle scattering amplitudes is shown to be in
disagreement with the classical result of General Relativity given by the
Schwarzschild solution. It is shown also that the potential so defined fails
to describe whatever non-Newtonian interactions of macroscopic bodies.
An alternative interpretation of the $\hbar^0$-order part of the loop
corrections is given directly in terms of the effective action.
Gauge independence of that part of the one-loop radiative corrections to
the gravitational form factors of the scalar particle is proved,
justifying the interpretation proposed.
\end{abstract}
PACS number(s): 04.60.Ds, 11.15.Kc, 11.10.Lm


\section{Introduction}

Apart from the issue of renormalization, quantization of the General
Theory of Relativity is carried out in much the same way as in the
case of ordinary Yang-Mills theories. Besides formalities of the consistent
quantization, however, there are questions of principle concerning
basic postulates underlying the synthesis of classical theory and
quantum-mechanical ideas. The rules of this synthesis are mainly
contained in the correspondence principle of N.Bohr, which, on the one hand,
gives recipe for the construction of operators for physical field quantities,
and, on the other hand, implies definite requirements as to the form of these
quantities in cases when a system displays classical properties.

Establishing the correspondence between the classical and quantum modes
of description in the case of the theory of gravity displays features
quite different from those encountered in other theories of fundamental
interactions.  Distinctions arise, in particular, when the limiting
procedure of transition from the quantum to classical theory is performed.
In theories like quantum electrodynamics, e.g., interaction of two charged
particles takes the form of the Coulomb law when momentum transfer from
one particle to the other becomes small as compared to the particles' masses,
so the above mentioned procedure is accomplished by tending the masses
to infinity. In essential, this constitutes most of what is called the
physical renormalization conditions. The latter require the gauge field
propagator (in the momentum space) to have the unit residue of the pole
at zero momentum, which is just the aforementioned condition on the form of
the two-particle interaction in the coordinate representation.

In the case of gravity, however, the situation is essentially different.
There, transition to the classical theory cannot be performed by taking
the limit: particle mass $\to \infty$, because the value of the particle mass
determines the strength of its gravitational interactions.
In fact, relative value of the radiative corrections to the classical
Newton law, corresponding to the logarithmical contribution to the
gravitational form factors of the scalar particle, is independent of the
scalar particle mass \cite{donoghue}.

Although this trait of the gravitational interaction makes it exceptional
among the others, and moreover, is in apparent contradiction with the
standard formulation of the correspondence principle, it is not
really an inconsistency in the quantum description of gravitation:
to get rid of it, one may simply weaken the correspondence principle,
and require the relative corrections to the Newton law to disappear
only at large distances between the particles.

There is, however, a still more important aspect of the correspondence
between the classical and quantum pictures of gravitation, that attracts
our attention in the present paper. The Einstein theory, being essentially
nonlinear, demands the quantum theory to reproduce not only the
Newtonian form of the particle interaction, but also all the nonlinear
corrections predicted by the General Relativity.
In this respect, the above-mentioned peculiarity of the gravitational
interaction, namely, its proportionality to the masses of particles,
is manifested in the fact (also pointed out in Ref.~\cite{donoghue})
that, along with the true quantum corrections (i.e., proportional to
the Planck constant $\hbar$), the loop contributions also contain
classical pieces (i.e., proportional to $\hbar^0$).
Thus, an important question arises as to relationship between these
classical loop contributions and the classical predictions
of the General Relativity.

In analogous situation in the Yang-Mills theories, the correct correspondence
between the classical and quantum pictures is guaranteed by the fact that all
the radiative corrections to the particle form factors disappear in the limit:
masses $\to \infty,$ thus providing the complete reduction of a given
quantum picture to the corresponding nonlinear classical solution.
It is claimed in Ref.~\cite{donoghue} that when collected in the course of
construction of the gravitational potential from the one-particle-reducible
Feynman graphs, aforesaid classical contributions just reproduce the
post-Newtonian terms given by the expansion of the Schwarzschild metric
in powers of $r_{g}/r,$ $r_{g}$ being the gravitational radius.

It will be shown in Sec.~\ref{definition} that this claim is erroneous:
the one-loop terms of the order $\hbar^0$ in the gravitational
potential are actually two times larger than the terms of the order
$r^2_{g}/r^2$ coming from the Schwarzschild expression. This in turn
raises the question of relevance of the notion of potential in the case of
quantum gravity. It will be shown in Sec.~\ref{extension} that not only the
value of the potential, defined with the help of the two-particle
scattering amplitudes, disagrees with the classical results of the General
Relativity, but also that the potential so constructed fails to describe
whatever non-Newtonian interaction of {\it macroscopic} bodies.
After that an alternative interpretation of the $\hbar^0$ parts of the loop
contributions will be suggested, based on a certain modification
of the correspondence principle. The use of the effective action formalism
turns out to be essential for the new interpretation, running thereby into
the problem of gauge dependence of the effective action. In Sec.~\ref{zero}, gauge
independence of the $\hbar^0$ part of the one-loop contribution to
the gravitational form factors of the scalar particle is proved, thus
ensuring the physical sense of our interpretation.
Sec.~\ref{tools} contains brief description of the method used in evaluation
of the gauge-dependent parts of the loop corrections.
The results of the work are discussed in Sec.~\ref{conclude}.
Some formulae needed in calculation of the Feynman integrals are obtained
in the Appendix.

We use the highly condensed notations of DeWitt \cite{dewitt1} throughout
this paper. Also left derivatives with respect to anticommuting variables
are used. The dimensional regularization of all divergent quantities
is supposed.

\section{Definition of the potential in quantum gravity}\label{definition}

Before we proceed to actual calculations, it is worthwhile to make
some remarks concerning the notion of potential in quantum theory.

Let us begin with an obvious but far-reaching observation that
a definition of potential in any (classical or quantum) field theory
must be given in terms characterizing motion of interacting particles,
simply because only in this case the definition would be relevant to
an experiment. For this purpose, the scattering matrix approach can
be used, in which case the potential is conventionally defined as the
Fourier transform (with respect to the momentum transfer from one particle
to the other) of the suitably normalized two-particle scattering amplitude.
By itself this definition is not of great value unless one is able to
separate the whole scattering process as follows: interaction of the
first particle with the gauge field $\to$ propagation of the gauge field
$\to$ interaction of the gauge field with the second particle.
Only if such a separation is possible can one introduce a self-contained
notion of the potential. In terms of the Feynman diagrams, one would say
in this case that the diagrams describing the scattering process are
one-particle-reducible with respect to the gauge field.

In connection with what just have been said, a question may arise of what
the construction of the potential, or some other object based on the
above-mentioned separation of the particle interaction,
is needed for. The answer is that only through
such a construction can the correspondence between the classical and quantum
modes of description be established. Indeed, the very nature of classical
conceptions implies existence of a self-contained notion of the field
produced by a given source, which value is independent of a specific
device chosen to measure it. An object possessing these properties would be
just supplied by the potential defined in the manner outlined
above.\footnote{In the case of essentially
non-linear theory such as the General Relativity, this potential will not be
proportional to the product of the charges of the scattering particles
(in the case of gravity -- to the product of their masses),
one of which plays the role of the source for the gauge field, and the other
-- the measuring device. However, this is not a problem, since one can always
imagine, again in the classical spirit, that the charge of the measuring
particle is small as compared with the charge of the source-particle.
Then the potential will be independent of this small charge.}

Definition of the potential through the scattering amplitudes is not the
only way to introduce an independent notion of the gauge field.
Till now, however, it {\it is} the only {\it consistent} way
if one is interested in giving a {\it gauge-independent}
definition, i.e., the one that would give values for the gauge field
independent of the choice of gauge conditions needed to fix gauge
invariance of the theory.\footnote{One also has to require independence of
the choice of a set of dynamical variables in terms of which the theory is
quantized. This last condition is particularly important in the case of
gravity, where one is free to take any tensor density as a dynamical
parametrization of the metric field.} Actually, it was recently proposed
that, in the case of quantum gravity, such a definition can be given beyond
the S-matrix approach through the introduction of classical point particle
moving in a given gravitational field and playing the role of a measuring
device \cite{dalvit}. In particular, it was shown that the one-loop effective
equations of motion of the point-particle (the effective geodesic equation),
calculated in the weak field approximation in the non-relativistic limit,
turn out to be independent of the gauge conditions fixing the general
covariance \cite{dalvit}. Although this result, undoubtedly, is of
considerable importance on its own, it lies out of the main line of our
concern here, since it is based on the introduction of the classical
point-particle into the functional integral "by hands", which certainly
cannot be justified using consistent limiting procedure of transition
from the underlying quantum field theory to the classical theory.
On the other hand, as was shown in Ref.~\cite{kazakov1}, introduction of
the classical {\it field} matter (scalar field) instead of the point-like
still leads to the gauge-dependent values for the gravitational
field.\footnote{It seems that in the case of ordinary Yang-Mills theories,
inclusion of the classical field matter does solve the gauge-dependence
problem, at least in the low-energy limit, see Ref.~\cite{kazakov2}.}

Therefore, it seems natural to try to establish the correspondence
between the quantum gravity and the classical General Relativity just
in terms of the potential defined through the scattering amplitudes.
However, it will be shown below that the value of the potential, found in
Ref.~\cite{donoghue}, disagrees with the classical result given by the
Schwarzschild solution. The latter has the form
\begin{eqnarray}\label{sch1}
ds^2 = \left(1-\frac{r_g}{r}\right) c^2 d t^2 - \frac{d r^2}{1-\frac{r_g}{r}}
- r^2 (d\theta^2 + \sin^2\theta\ d\varphi^2),
\end{eqnarray}
\noindent
where $\theta, \varphi$ are the standard spherical angles, $r$ is the radial
coordinate, and $r_g = 2 G M/c^2$ is the gravitational radius of a
spherically-symmetric distribution of mass $M.$ The form of $d s^2$ given
by Eq.~(\ref{sch1}) is fixed by the requirements $g_{ti} = 0,$
$i = r, \theta, \varphi,$ $g_{\theta\theta} = r^2.$ To compare the two
results, however, one has to transform Eq.~(\ref{sch1}) to the DeWitt gauge
\begin{eqnarray}\label{gauge}
\eta^{\mu\nu}\partial_{\mu} g_{\nu\alpha}
- \frac{1}{2}\eta^{\mu\nu}\partial_{\alpha} g_{\mu\nu} = 0,
\end{eqnarray}
\noindent
used in Ref.~\cite{donoghue}.

The $t, \theta, \varphi$-components of Eq.~(\ref{gauge}) are already
satisfied by the solution (\ref{sch1}). To meet the remaining condition,
let us substitute $r \to f(r),$ where $f$ is a function of $r$ only.
Then the $t, \theta, \varphi$-components of Eq.~(\ref{gauge}) are still
satisfied, while its $r$-component gives the following equation for the
function $f(r):$
\begin{eqnarray}\label{rcond}
\frac{1}{r^2}\frac{\partial}{\partial r}\left(\frac{r^2 f'^2}{1-r_g/f}\right)
-\frac{2 f^2}{r^2}
- \frac{1}{2}\frac{\partial}{\partial r}\left(1-\frac{r_g}{f}
+ \frac{f'^2}{1-r_g/f}+\frac{2 f^2}{r^2}\right) = 0,
\end{eqnarray}
\noindent
where $f'\equiv \partial f(r)/\partial r.$

Since we are interested only in the long-distance corrections to
the Newton law, one may expand $f(r)/r$ in powers of $r_g/r$ keeping
only the first few terms:
$$f(r) = r\left[1 + c_1 \frac{r_g}{r} + c_2 \left(\frac{r_g}{r}\right)^2
+ \cdot\cdot\cdot \right].$$ Substituting this into Eq.~(\ref{rcond}),
one obtains successively $c_1 =1/2,$ $c_2=1/2,$ etc.

Therefore, up to terms of the order $r^2_g/r^2,$ the Schwarzschild solution
takes the following form
\begin{eqnarray}\label{sch2}&&
ds^2 = \left(1 - \frac{r_g}{r} + \frac{r^2_g}{2 r^2}\right) c^2 d t^2
- \left(1 + \frac{r_g}{r} - \frac{r^2_g}{2 r^2}\right) d r^2
\nonumber\\&&
- r^2 \left(1 + \frac{r_g}{r} + \frac{5 r^2_g}{4 r^2}\right) (d\theta^2 + \sin^2\theta\ d\varphi^2).
\end{eqnarray}
Taking the square root of the time component of the metric (\ref{sch2}),
we see that the classical gravitational potential turns out to be equal to
\begin{eqnarray}\label{sch3}&&
\Phi^c(r) =  - \frac{G M}{r} + \frac{G^2 M^2}{2 c^2 r^2}.
\end{eqnarray}
The post-Newtonian correction is here two times smaller than that obtained
in \cite{donoghue}.

Thus, we arrive at the puzzling conclusion that in the classical limit,
the quantum theory of gravity, being based on the Bohr correspondence
principle, does not reproduce the Einstein theory it originates from.
However, before making such a conclusion, one would question relevance of
the notion of the gravitational potential itself.
It may well turn out that this discrepancy arises because of incorrect
choice of the quantum-field quantity to be traced back to the classical
potential. That this is very likely so will be demonstrated in the
sections below.

\section{Classical loop corrections and the correspondence principle}\label{extension}

As was explained in Sec.~\ref{definition}, establishing correspondence
between the quantum and classical theories of gravitation turns out
to be highly nontrivial in view of the fact that the classical contribution
to a given process is not contained entirely in the trees, but comes also
from the loop diagrams. It is not even clear what the quantum-theoretical
counterpart of the classical potential (or, more generally, of the classical
metric) is. As we saw in Sec.~\ref{definition}, the usual definition of the
quantum gravitational potential is not in concord with the classical theory.

It is worthwhile to emphasize that our consideration is not restricted
to the case of the Einstein theory only. Being related to the low-energy
phenomena, all the conclusions are valid in any quantum theory of gravity
as well, if that theory becomes Newtonian in the non-relativistic limit.

There are different ways of thinking in this situation. One could conclude,
for example, that since the quantum gravity does not satisfy the
correspondence principle, the basic postulates of the General Relativity are
incompatible with the principles of quantum theory, and therefore their
synthesis is impossible. Or, one could try to find another definition
of the gravitational potential, that would be in agreement with the
classical General Relativity, on the one hand, and possessed physical
meaning at the quantum level, on the other.

Below, a more refined interpretation will be given, based on a certain
modification of the correspondence principle.

\begin{figure}
\epsfxsize=15cm\epsfbox{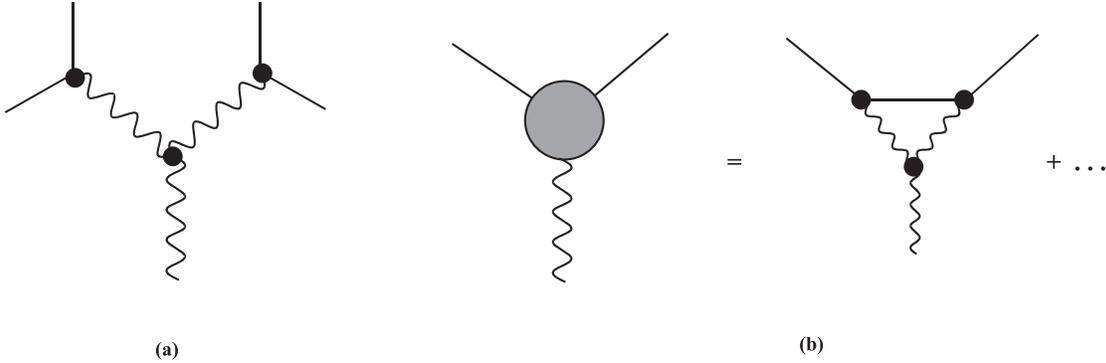}
\vspace{1cm}
\caption{Diagrams contributing to the first post-Newtonian correction.
(a) The tree diagram occurring because of the nonlinearity of the
Einstein equations. (b) The one-loop form factor.  Wavy lines represent
gravitons, solid lines scalar particles.}
\label{fig1}
\end{figure}

Let us first note that, from the formal point of view, the correspondence
between any quantum theory and its classical original is most naturally
established in terms of the effective action rather than the S-matrix.
This is because the effective action [the generating functional of
the one-particle-irreducible Green functions, $\Gamma,$ defined
in Eq.~(\ref{effaction}) below] just coincides with the
initial classical action in the tree approximation. In particular,
nonlinearity of the classical theory (resulting, e.g., in $r_g/r$-
power corrections to the Newton law in the General Relativity) is
correctly reproduced by the trees. In the case of quantum gravity,
there are still additional contributions of the order $\hbar^0,$ coming from
the loop corrections. They are given just by the gravitational form factors
of particles, which serve as the building blocks for the gravitational
potential. Instead of constructing the potential, however, let us consider
them in the framework of the effective action method. From the point of view
of this method, the $\hbar^0$-parts of the particle form factors, together
with the proper quantum parts of the order $\hbar,$ describe the
radiative corrections to the classical equations of motion of the
gravitational field. It is non-vanishing of these terms that violates
the usual Bohr correspondence. There is, however, one essential difference
between the $\hbar^0$ loop terms and the nonlinear tree corrections.
Consider, for instance, the first post-Newtonian correction of the form
$${\rm const} \frac{G^2 M^2}{c^2 r^2}.$$ The "const" receives contributions
from the tree diagram pictured in Fig.~\ref{fig1}(a), as well as from the
one-loop form factor, Fig.~\ref{fig1}(b). If the gravitational field is
produced by only one particle of mass $M,$ then the two contributions
are of the same order of magnitude.

They are not, however, if the field is produced by a
macroscopic body consisting of a large number $N$ of particles with mass
$m = M/N.$ Being responsible for the nonlinearity of Einstein equations,
the tree diagram \ref{fig1}(a) is {\it bilinear} in the energy-momentum tensor
$T^{\mu\nu}$ of the particles, while the loop diagram \ref{fig1}(b) is only
{\it linear} (to be precise, it has only two particle operators attached).
Therefore, when evaluated between $N$-particle state, the former is
proportional to $(m \cdot N) \cdot (m\cdot N) = M^2,$ while the latter --
to $m^2 \cdot N = M^2/N.$ If, for instance, the solar gravitational field
is considered, the quantum correction is suppressed by a factor of the order
$m_{proton}/M_{\odot} \approx 10^{-57}.$

This fact suggests the following interpretation of the correspondence
principle when applied to the case of gravity:
{\it the effective gravitational field
produced by a macroscopic body of mass $M$ consisting of $N$ particles
turns into corresponding classical solution of the Einstein equations
in the limit $M \to \infty,$ $N \to \infty.$ }

It is clear from the above discussion that the use of the effective action
is essential for this interpretation:
the quantum potential, being of the form
$$\Phi^{q}(r) = - N \frac{G (M/N)}{r}
+ {\rm const}_1 N \frac{G^2 (M/N)^2}{c^2 r^2}
+ {\rm const}_2 N \frac{G^3 (M/N)^3}{c^4 r^3} + \cdot\cdot\cdot,$$
for a macroscopic ($N$-particle) body of mass $M,$ would fail to reproduce
classical potential other than Newtonian, since
$$\Phi^{q}(r) \to - \frac{G M}{r} ~~{\rm when} ~~N \to \infty.$$

Thus, the loop corrections of the order $\hbar^0$ are now considered
on a equal footing with the tree corrections, and thereby are endowed direct
{\it physical meaning as describing deviations of the space-time metric from
classical solutions of the Einstein equations in the case of finite $N.$ }

Like any other argument trying to assign physical meaning to the effective
action, the above interpretation immediately runs up against the problem of
its gauge dependence. In spite of being independent of the Planck constant,
the $\hbar^0$ terms originating from the loop diagrams are not
gauge-independent {\it a priori}. However, as will be demonstrated below,
there is a strong evidence for that they {\it are} gauge-independent
nevertheless. Namely, the $\hbar^0$ terms of the gravitational form factors
of the scalar particle, contributing to the first post-Newtonian correction
to the metric, turn out to be gauge-independent.\footnote{To be precise,
one has to speak about gauge dependence of scalar quantities, such as the
scalar curvature, built out of the metric, rather than the metric itself,
since the latter is gauge-dependent by definition (Cf., for instance,
Eqs.~(\ref{sch1}),~(\ref{sch2}) representing the Schwarzschild solution
for two different gauge conditions).}

\section{Generating functionals and Slavnov identities}\label{tools}

As in Ref.~\cite{donoghue}, we consider the system of quantized
gravitational and scalar matter fields. Dynamics of the scalar field
denoted by $\phi,$ is described by the action
\begin{eqnarray}&&\label{actionm}
S_{\phi} =  \frac{1}{2}{\displaystyle\int} d^4 x \sqrt{-g}(g^{\mu\nu}\partial_{\mu}\phi \partial_{\nu}\phi - m^2 \phi^2),
\end{eqnarray}
\noindent
while the action for the gravitational field\footnote{Our notation
is $R_{\mu\nu} \equiv R^{\alpha}_{~\mu\alpha\nu} =
\partial_{\alpha}\Gamma^{\alpha}_{\mu\nu} - \cdot\cdot\cdot,
~R \equiv R_{\mu\nu} g^{\mu\nu}, ~g\equiv \det g_{\mu\nu},
~g_{\mu\nu} = {\rm sgn}(+,-,-,-).$
Dynamical variables of the gravitational field
$h_{\mu\nu} = g_{\mu\nu} - \eta_{\mu\nu},
\eta_{\mu\nu} = {\rm diag}\{+1,-1,-1,-1\}.$}
\begin{eqnarray}&&\label{actionh}
S = - \frac{1}{k^2}{\displaystyle\int} d^4 x \sqrt{-g}R,
\end{eqnarray}
$k$ being the gravitational constant.\footnote{We choose units in
which $c = \hbar = k = 1$ from now on.}

The action $S + S_{\phi}$ is invariant under the following
gauge transformations\footnote{Indices of the
functions $F, \xi$, as well as of the ghost fields below,
are raised and lowered, if convenient, with the help of Minkowski metric $\eta_{\mu\nu}$.}
\begin{eqnarray}&&\label{gaugesym}
\delta h_{\mu\nu} = \xi^{\alpha}\partial_{\alpha}h_{\mu\nu}
+ (\eta_{\mu\alpha} + h_{\mu\alpha})\partial_{\nu}\xi^{\alpha}
+ (\eta_{\nu\alpha} + h_{\nu\alpha})\partial_{\mu}\xi^{\alpha}
\equiv D_{\mu\nu}^{\alpha}(h)\xi_{\alpha},
\nonumber\\&&
~~\delta\phi = \xi^{\alpha}\partial_{\alpha}\phi \equiv
\tilde{D}^{\alpha}(\phi)\xi_{\alpha},
\end{eqnarray}
where $\xi^{\alpha}$ are the (infinitesimal) gauge functions.
The generators $D,\tilde{D}$ span the closed algebra
\begin{eqnarray}&&\label{algebra}
D_{\mu\nu}^{\alpha,\sigma\lambda} D_{\sigma\lambda}^{\beta}
- D_{\mu\nu}^{\beta,\sigma\lambda} D_{\sigma\lambda}^{\alpha}
= f_{~~~\gamma}^{\alpha\beta} D_{\mu\nu}^{\gamma},
\nonumber\\&&
\tilde{D}^{\alpha}_{1} \tilde{D}^{\beta}
- \tilde{D}^{\beta}_1 \tilde{D}^{\alpha} = f^{\alpha\beta}_{~~~\gamma} \tilde{D}^{\gamma},
\end{eqnarray}
the "structure constants" $f^{\alpha\beta}_{~~~\gamma}$ being defined by
\begin{eqnarray}&&
f_{~~~\gamma}^{\alpha\beta}\xi_{\alpha}\eta_{\beta} =
\xi_{\alpha}\partial^{\alpha}\eta_{\gamma}
- \eta_{\alpha}\partial^{\alpha}\xi_{\gamma}.
\end{eqnarray}

Let the gauge invariance be fixed by the term
\begin{eqnarray}\label{gaugefix}&&
S_{gf} = \frac{1}{2\xi}\eta^{\alpha\beta} F_{\alpha} F_{\beta},
\nonumber\\&&
~~F_{\alpha} = \partial^{\mu} h_{\mu\alpha} - \frac{1}{2}\partial_{\alpha} h,
~~h \equiv \eta^{\mu\nu} h_{\mu\nu}.
\end{eqnarray}

Next, introducing the Faddeev-Popov ghost fields
$C_{\alpha}, \bar{C}^{\alpha}$ we write the Faddeev-Popov quantum action
\cite{faddeev}
\begin{eqnarray}\label{fp}
S_{FP} = S + S_{\phi} + S_{gf}
+ \bar{C}^{\beta}F_{\beta}^{,\mu\nu}D_{\mu\nu}^{\alpha}C_{\alpha}.
\end{eqnarray}
$S_{FP}$ is still invariant under the following BRST transformations \cite{brst}
\begin{eqnarray}\label{brst}&&
\delta h_{\mu\nu} = D_{\mu\nu}^{\alpha}(h)C_{\alpha}\lambda,
\nonumber\\&&
\delta \phi = \tilde{D}^{\alpha}(\phi)C_{\alpha}\lambda,
\nonumber\\&&
\delta C_{\gamma} = - \frac{1}{2}f^{\alpha\beta}_{~~~\gamma}C_{\alpha}C_{\beta}\lambda,
\nonumber\\&&
\delta \bar{C}^{\alpha} = \frac{1}{\xi}F^{\alpha}\lambda,
\end{eqnarray}
$\lambda$ being a constant anticommuting parameter.

The generating functional of Green functions\footnote{For brevity, the product symbol,
as well as tensor indices of the fields $h_{\mu\nu},
C_{\alpha}, \bar{C}^{\alpha},$ is omitted in the path integral measure.}
\begin{eqnarray}\label{gener}&&
Z[T,J,\bar{\beta},\beta,K,\tilde{K},L]
\nonumber\\&&
= {\displaystyle\int}dh d\phi dC d\bar{C} \exp\{i (\Sigma
+ \bar{\beta}^{\alpha}C_{\alpha} + \bar{C}^{\alpha}\beta_{\alpha} + T^{\mu\nu}h_{\mu\nu} + J\phi )\},
\end{eqnarray}
\noindent
where
\begin{eqnarray}&&
\Sigma = S_{FP}
+ K^{\mu\nu}D_{\mu\nu}^{\alpha}C_{\alpha}
+ \tilde{K}\tilde{D}^{\alpha}C_{\alpha}
+ L^{\gamma} \frac{1}{2} f^{\alpha\beta}_{~~~\gamma}C_{\alpha}C_{\beta},
\nonumber
\end{eqnarray}
the sets \{$T^{\mu\nu},$ $J,$ $\bar{\beta}^{\alpha},$ $\beta_{\alpha}$\} and
\{$K^{\mu\nu},$ $\tilde{K}$, $L^{\alpha}$\} being the sources for the fields
and BRST transformations, respectively \cite{zinnjustin}.

To determine the dependence of field-theoretical quantities on the gauge
parameter $\xi$, we modify the quantum action adding the term
\begin{eqnarray}\label{yform}
Y F_{\alpha}\bar{C}^{\alpha},
\nonumber
\end{eqnarray}
$Y$ being a constant anticommuting parameter \cite{nielsen}.
Thus we write the generating functional of Green functions as
\begin{eqnarray}\label{genernew}&&
Z[T,J,\bar{\beta},\beta,K,\tilde{K},L,Y]
= {\displaystyle\int}dh d\phi dC d\bar{C} \exp\{i (\Sigma
\nonumber\\&&
+ Y F_{\alpha}\bar{C}^{\alpha}
+ \bar{\beta}^{\alpha}C_{\alpha} + \bar{C}^{\alpha}\beta_{\alpha} + T^{\mu\nu}h_{\mu\nu} + J\phi )\}.
\end{eqnarray}

Finally, we introduce the generating functional of
connected Green functions
\begin{eqnarray}\label{defw}
W[T,J,\bar{\beta},\beta,K,\tilde{K},L,Y]=
- i \ln Z[T,J,\bar{\beta},\beta,K,\tilde{K},L,Y],
\end{eqnarray}
and then define the effective action $\Gamma$ in the usual way
as the Legendre transform of $W$ with respect to the mean fields
\begin{eqnarray}&&\label{mean}
h_{\mu\nu} = \frac{\delta W}{\delta T^{\mu\nu}},
~~\phi = \frac{\delta W}{\delta J},
~~C_{\alpha} = \frac{\delta W}{\delta\bar{\beta}^{\alpha}},
~~\bar{C}^{\alpha} = - \frac{\delta W}{\delta\beta_{\alpha}},
\end{eqnarray}
(denoted by the same symbols as the corresponding field operators):
\begin{eqnarray}&&\label{effaction}
\Gamma[h,\phi,C,\bar{C},K,\tilde{K},L,Y]
\nonumber\\&&
= W [T,J,\bar{\beta},\beta,K,\tilde{K},L,Y]
-  \bar{\beta}^{\alpha}C_{\alpha} - \bar{C}^{\alpha}\beta_{\alpha} - T^{\mu\nu}h_{\mu\nu} - J\phi.
\end{eqnarray}

Evaluation of derivatives of diagrams with respect to the gauge parameters
is a more easy task than their direct calculation in arbitrary
gauge.\footnote{In actual quantum gravity calculations, this fact was
first used in \cite{tutin} to evaluate divergences of the Einstein gravity
in arbitrary gauge off the mass shell.} This is because these derivatives
can be expressed through another set of diagrams with more simple structure.
The rules for such a transformation of diagrams are conveniently summarized
in the Slavnov identities corresponding to the generating functional
(\ref{genernew}). Since these identities are widely used in what follows,
their derivation will be briefly described below \cite{nielsen}.

First of all, we perform a BRST shift (\ref{brst}) of integration
variables in the path integral (\ref{genernew}). Equating the variation
to zero we obtain the following identity
\begin{eqnarray}&&\label{slav}
{\displaystyle\int}dh d\phi dC d\bar{C}
\left[i Y \bar{C}^{\alpha} F_{\alpha}^{,\mu\nu} D^{\beta}_{\mu\nu} C_{\beta}
+ i \frac{Y}{\xi} F_{\alpha}^{2}
+ T^{\mu\nu} \frac{\delta}{\delta K^{\mu\nu}}
+ J \frac{\delta}{\delta\tilde{K}}
- \bar{\beta}^{\alpha}\frac{\delta}{\delta L^{\alpha}}
- i \beta_{\alpha}\frac{F^{\alpha}}{\xi}
\right]
\nonumber\\&&
\exp\{i (\Sigma
+ Y F_{\alpha}\bar{C}^{\alpha}
+ \bar{\beta}^{\alpha}C_{\alpha} + \bar{C}^{\alpha}\beta_{\alpha}
+ T^{\mu\nu}h_{\mu\nu} + J\phi)\} = 0.
\end{eqnarray}
\noindent
Next, the second term in the square brackets in Eq.~(\ref{slav})
can be transformed with the help of the quantum ghost equation of motion,
obtained by performing a shift $\bar{C} \to \bar{C} + \delta\bar{C}$
of integration variables in the functional integral (\ref{genernew}):
\begin{eqnarray}\label{ghosteq}&&
{\displaystyle\int}dh d\phi dC d\bar{C}
\left[F_{\gamma}^{,\mu\nu}D_{\mu\nu}^{\alpha}C_{\alpha}
- Y F_{\gamma} + \beta_{\gamma} \right]
\exp\{\cdot\cdot\cdot\} = 0,
\nonumber
\end{eqnarray}
\noindent
from which it follows that
\begin{eqnarray}\label{ghosteq1}&&
Y {\displaystyle\int}dh d\phi dC d\bar{C}
\left[i \bar{C}^{\gamma}F_{\gamma}^{,\mu\nu}D_{\mu\nu}^{\alpha}C_{\alpha}
+ \beta_{\gamma}\frac{\delta}{\delta\beta_{\gamma}}\right]
\exp\{\cdot\cdot\cdot\} = 0,
\nonumber
\end{eqnarray}
\noindent
where we used the property $Y^2 = 0$, and omitted the expression
$\delta\beta_{\gamma}/\delta\beta_{\gamma} \sim \delta(0)$.
Putting this all together, we rewrite Eq.~(\ref{slav})
\begin{eqnarray}\label{slav1}
\left( T^{\mu\nu}\frac{\delta}{\delta K^{\mu\nu}}
+ J\frac{\delta}{\delta \tilde{K}}
- \bar{\beta}^{\alpha}\frac{\delta}{\delta L^{\alpha}}
- \frac{1}{\xi} \beta_{\alpha}F^{\alpha,\mu\nu}\frac{\delta}{\delta T^{\mu\nu}}
- Y \beta_{\gamma}\frac{\delta}{\delta\beta_{\gamma}}
- 2 Y\xi\frac{\partial}{\partial\xi}
\right) Z  = 0.
\end{eqnarray}
This is the Slavnov identity for the generating functional of Green functions
we are looking for.
In terms of the generating functional of connected Green functions,
it looks like
\begin{eqnarray}\label{slav2}
T^{\mu\nu}\frac{\delta W}{\delta K^{\mu\nu}}
+ J\frac{\delta W}{\delta \tilde{K}}
- \bar{\beta}^{\alpha}\frac{\delta W}{\delta L^{\alpha}}
- \frac{1}{\xi} \beta_{\alpha}F^{\alpha,\mu\nu}\frac{\delta W}{\delta T^{\mu\nu}}
- Y \beta_{\gamma}\frac{\delta W}{\delta\beta_{\gamma}}
- 2 Y\xi\frac{\partial W}{\partial\xi} = 0.
\end{eqnarray}
\noindent
It can be transformed further into an identity for the generating functional
of proper vertices: with the help of equations
\begin{eqnarray}&&\label{meaninv}
T^{\mu\nu} =  - \frac{\delta \Gamma}{\delta h_{\mu\nu}},
~~J =  - \frac{\delta \Gamma}{\delta \phi},
~~\bar{\beta}^{\alpha} = \frac{\delta \Gamma}{\delta C_{\alpha}},
~~\beta_{\alpha} = - \frac{\delta \Gamma}{\delta\bar{C}^{\alpha}},
\nonumber
\end{eqnarray}
which are the inverse of Eqs.\ (\ref{mean}),
and the relations
\begin{eqnarray}
\frac{\delta W}{\delta K^{\mu\nu}} = \frac{\delta \Gamma}{\delta K^{\mu\nu}},
~~\frac{\delta W}{\delta \xi} = \frac{\delta \Gamma}{\delta \xi}, {\rm ~~etc.}
\nonumber
\end{eqnarray}
we rewrite Eq.~(\ref{slav2})
\begin{eqnarray}\label{slav3}&&
\frac{\delta \Gamma}{\delta h_{\mu\nu}}\frac{\delta \Gamma}{\delta K^{\mu\nu}}
+ \frac{\delta \Gamma}{\delta \phi}\frac{\delta \Gamma}{\delta \tilde{K}}
+ \frac{\delta \Gamma}{\delta C_{\alpha}}\frac{\delta \Gamma}{\delta L^{\alpha}}
- \frac{F^{\alpha}}{\xi} \frac{\delta \Gamma}{\delta\bar{C}^{\alpha}}
+ Y \frac{\delta \Gamma}{\delta\bar{C}^{\alpha}}\bar{C}^{\alpha}
+ 2 Y\xi\frac{\partial \Gamma}{\partial\xi} = 0.
\end{eqnarray}
\noindent
Written down via the reduced functional
\begin{eqnarray}\label{checkg}&&
{\Gamma_0} = \Gamma - \frac{1}{2\xi}F_{\alpha} F^{\alpha}
- Y F_{\sigma}\bar{C}^{\sigma},
\end{eqnarray}
the latter equation takes particularly simple form
\begin{eqnarray}\label{slav4}&&
\frac{\delta\Gamma_0}{\delta h_{\mu\nu}}\frac{\delta\Gamma_0}{\delta K^{\mu\nu}}
+ \frac{\delta\Gamma_0}{\delta \phi}\frac{\delta\Gamma_0}{\delta \tilde{K}}
+ \frac{\delta\Gamma_0}{\delta C_{\sigma}}\frac{\delta\Gamma_0}{\delta L^{\sigma}}
+ 2 Y \xi\frac{\partial\Gamma_0}{\partial\xi} = 0.
\end{eqnarray}

\section{The $\hbar^0$ radiative corrections to the gravitational form factors}\label{zero}

Let us now turn to the explicit evaluation of the radiative corrections.
In this section, $\xi$-independence of the loop amendment to the first
post-Newtonian classical correction given by the second term in
Eq.~(\ref{sch3}) will be proved. The only loop diagram we need to
consider is the one pictured in Fig.~\ref{fig1}(b). Other one-loop diagrams
do not contain terms proportional to the $\sqrt{- p^2},$ responsible
for the $1/r^2$ behavior of the form factors, while higher-loop
diagrams are of higher orders in the Newton constant $G.$

To evaluate the $\xi$-derivative of diagram \ref{fig1}(b), we use the Slavnov
identity (\ref{slav4}). Extracting terms proportional to the source $Y,$
we get
\begin{eqnarray}\label{slav4y}&&
2\xi\frac{\partial\Gamma_{1}}{\partial\xi} =
 \frac{\delta\Gamma_{1}}{\delta h_{\mu\nu}}\frac{\delta\Gamma_{2}}{\delta K^{\mu\nu}}
+ \frac{\delta\Gamma_{1}}{\delta \phi}\frac{\delta\Gamma_{2}}{\delta \tilde{K}},
\end{eqnarray}
\noindent
where $\Gamma_{1,2}$ are defined by
$$\Gamma_{1} = \Gamma_{0}|_{Y=0}, ~~\Gamma_{2}
= \frac{\partial\Gamma_{0}}{\partial Y}.$$

\begin{figure}
\epsfxsize=15cm\epsfbox{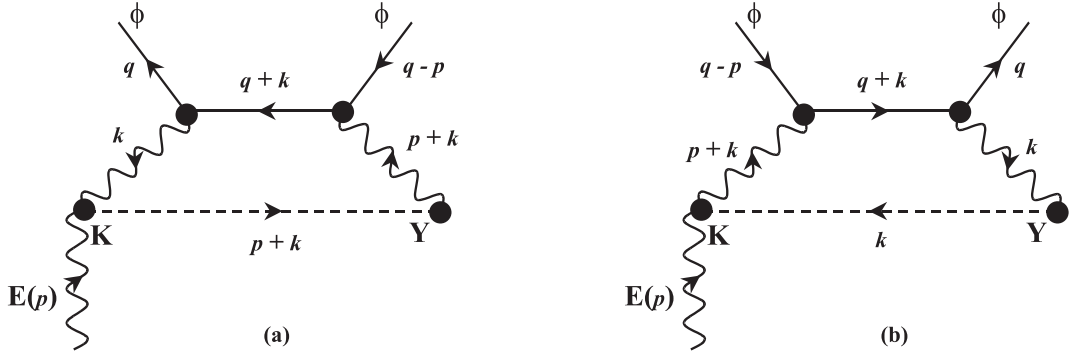}
\vspace{1cm}
\caption{Diagrams with two scalar and one graviton external lines,
containing the terms proportional to the square root of $(-p^2)$
in the right hand side of Eq.~(26). Dashed lines represent ghosts.}
\label{fig2}
\end{figure}

At the one-loop level, Eq.~(\ref{slav4y}) is just\footnote{Enclosed
in the round brackets is the number of loops in the diagram representing
a given term.}
\begin{eqnarray}\label{slav4y1}&&
2\xi\frac{\partial\Gamma^{(1)}_{1}}{\partial\xi} =
\frac{\delta \Gamma^{(0)}_{1}}{\delta h_{\mu\nu}}\frac{\delta\Gamma^{(1)}_{2}}{\delta K^{\mu\nu}},
\end{eqnarray}
since the external scalar lines are on the mass shell
$$\frac{\delta S_{\phi}}{\delta \phi} = 0.$$
\noindent
Diagrams with two scalar and one graviton external lines, giving
rise to the root singularity in the right hand side of Eq.~(\ref{slav4y1}),
are represented in Fig.~\ref{fig2}.

The corresponding analytical expressions\footnote{For simplicity, the
normalization factors coming from the external scalar lines, are omitted.}
\begin{eqnarray}&&\label{int}
I_{3(a)}(p,q) = - i E^{\mu\nu}(p) ~\mu^{\varepsilon} {\displaystyle\int} \frac{d^{4-\varepsilon} k}{(2\pi)^4}
\left\{
\frac{1}{2} W^{\alpha\beta\gamma\delta} q_{\gamma}(k_{\delta} + q_{\delta})
- m^2 \frac{\eta^{\alpha\beta}}{2}
\right\} G_{\phi}
\nonumber\\&&
\times\left\{
\frac{1}{2} W^{\rho\tau\sigma\lambda}(q_{\sigma} - p_{\sigma} )
(k_{\lambda} + q_{\lambda}) - m^2 \frac{\eta^{\rho\tau}}{2}
\right\}
\xi D^{(0)\eta}_{\rho\tau}\tilde{G}_{\eta}^{\xi}(k+p),
\nonumber\\&&
\times\tilde{G}_{\xi}^{\zeta}(p+k)
\left\{k_{\zeta} \delta_{\mu\nu}^{\chi\theta} -
\delta_{\zeta\mu}^{\chi\theta} (k_{\nu} + p_{\nu}) -
\delta_{\zeta\nu}^{\chi\theta} (k_{\mu} + p_{\mu})\right\}
G_{\chi\theta\alpha\beta}(k),
\end{eqnarray}
$$I_{3(b)}(p,q) = I_{3(a)}(p,p-q),$$
\noindent
where the following notation is introduced:
\begin{eqnarray}
W^{\alpha\beta\gamma\delta} =
\eta^{\alpha\beta} \eta^{\gamma\delta}
- \eta^{\alpha\gamma} \eta^{\beta\delta}
- \eta^{\alpha\delta} \eta^{\beta\gamma},
~~\delta_{\alpha\beta}^{\mu\nu} = \delta_{\alpha}^{\mu}\delta_{\beta}^{\nu}
+ \delta_{\alpha}^{\nu}\delta_{\beta}^{\mu},
\nonumber
\end{eqnarray}
\noindent
$G_{\mu\nu\sigma\lambda}$ is the graviton propagator defined by
$$\frac{\delta^2 S}{\delta h_{\rho\tau}\delta h_{\mu\nu}}
G_{\mu\nu\sigma\lambda} = - \delta_{\sigma\lambda}^{\rho\tau},$$
\begin{eqnarray}&&\label{pr}
G_{\mu\nu\sigma\lambda} =
- \frac{W_{\mu\nu\sigma\lambda}}{\Box}
+ (\xi - 1) (\eta_{\mu\sigma} \partial_{\nu} \partial_{\lambda}
+ \eta_{\mu\lambda} \partial_{\nu} \partial_{\sigma}
+ \eta_{\nu\sigma} \partial_{\mu} \partial_{\lambda}
+ \eta_{\nu\lambda} \partial_{\mu} \partial_{\sigma}) \frac{1}{\Box^2},
\end{eqnarray}
\noindent
$\tilde{G}^{\alpha}_{\beta}$ is the ghost propagator
\begin{eqnarray}&&\label{gt1}
\tilde{G}^{\alpha}_{\beta} = - \frac{\delta^{\alpha}_{\beta}}{\Box},
\nonumber
\end{eqnarray}
satisfying
\begin{eqnarray}&&
F_{\alpha}^{,\mu\nu}D^{(0)\beta}_{\mu\nu}\tilde{G}^{\gamma}_{\beta}
= - \delta_{\alpha}^{\gamma},
~~D^{(0)\alpha}_{\mu\nu} \equiv D^{\alpha}_{\mu\nu}(h=0),
\nonumber
\end{eqnarray}
\noindent
$G_{\phi}$ is the scalar particle propagator
$$G_{\phi} =  \frac{1}{\Box + m^2},$$
\noindent
$E^{\mu\nu}$ stands for the linearized Einstein tensor
\begin{eqnarray}
E^{\mu\nu} = R^{\mu\nu} - \frac{1}{2}\eta^{\mu\nu} R_{\alpha\beta} \eta^{\alpha\beta},
~~R_{\mu\nu} = \frac{1}{2}(\partial^{\alpha}\partial_{\mu} h_{\alpha\nu}
+ \partial^{\alpha}\partial_{\nu} h_{\alpha\mu}
- \Box h_{\mu\nu} - \partial_{\mu}\partial_{\nu} h),
\nonumber
\end{eqnarray}
\noindent
$\mu$ -- arbitrary mass scale, and $\varepsilon = 4 - d,$ $d$ being the
dimensionality of space-time.
\noindent
To simplify the tensor structure of diagram Fig.~\ref{fig2}(a), the use
has been made of the identity
\begin{eqnarray}&&
\frac{1}{\xi}F^{\alpha,\mu\nu} G_{\mu\nu\sigma\lambda}(x) =
- D^{(0)\beta}_{\sigma\lambda}\tilde{G}_{\beta}^{\alpha}(x),
\nonumber
\end{eqnarray}
which is nothing but the well-known {\it first} Slavnov identity at the
tree level; it is easily obtained differentiating Eq.~(\ref{slav2}) twice
with respect to $\beta_{\alpha}$ and $T^{\mu\nu}$, and setting all the
sources equal to zero.

The tensor multiplication in Eq.~(\ref{int}) is conveniently performed
with the help of the new tensor package for the REDUCE system \cite{reduce}
\begin{eqnarray}&&\label{int1}
I_{3(a)}(p,q) = - i E^{\mu\nu}(p) ~\mu^{\varepsilon} {\displaystyle\int}
\frac{d^{4-\varepsilon} k}{(2\pi)^4} \frac{1}{k^4}\frac{1}{(k+p)^4}
\frac{1}{m^2-(k+q)^2}
\nonumber\\&&
\times \xi [
\eta_{\mu\nu} k^2 m^2 \{(kp) - (kq)\}\{k^2 + 2 (kq)\}
+ k_{\mu} k_{\nu} k^4 \xi (p^2 - 2 m^2)
\nonumber\\&&
+ 2 k_{\mu} k_{\nu} k^2 (kq) (2 \xi - 1) (p^2 - 2 m^2)
+ 4 k_{\mu} (k_{\nu} + p_{\nu}) (kq)^2 (\xi - 1) (p^2 - 2 m^2)
\nonumber\\&&
+ k_{\mu} p_{\nu} k^4 (- 2 \xi m^2 + \xi p^2 - 2 m^2)
+ 2 k_{\mu} p_{\nu} k^2 (kq)(- 4 \xi m^2 + 2 \xi p^2 - p^2)
\nonumber\\&&
+  2 k_{\mu} q_{\nu} k^4 (p^2 - m^2)
+  4 k_{\mu} q_{\nu} k^2 (kq) (p^2 - m^2)
- 2 p_{\mu} p_{\nu} k^2 m^2 \{k^2 + 2 (kq)\}
\nonumber\\&&
+ 2 p_{\mu} q_{\nu} k^2 \xi \{(kp) - (kq)\} \{k^2 + 4 (kq)\}
+ 4 p_{\mu} q_{\nu} k^2 (kq) \{ p^2 - m^2 + (kq) - (kp)\}
\nonumber\\&&
+ 8 p_{\mu} q_{\nu} (kq)^2 (\xi-1)\{(kp) - (kq)\}
+ 2 p_{\mu} q_{\nu} k^4 (p^2 - m^2)
\nonumber\\&&
+ 2 q_{\mu} q_{\nu} k^4 \{(kq)- (kp)\}
+ 4 q_{\mu} q_{\nu} k^2 (kq) \{(kq) - (kp)\}
].
\end{eqnarray}
\noindent
Evaluation of the loop integrals can be automatized to a considerable
extent if the Schwinger parametrization of denominators
in Eq.~(\ref{int1}) is used
\begin{eqnarray}&&
\frac{1}{k^4} = \int_{0}^{\infty} dy~y \exp\{y k^2\},
~~\frac{1}{(k + p)^4} = \int_{0}^{\infty} dx~x \exp\{x (k + p)^2\},
\nonumber\\&&
\frac{1}{k^2 + 2 (kq)} = - \int_{0}^{\infty} dz \exp\{z [k^2 + 2 (kq)]\}.
\end{eqnarray}
It is convenient to apply these formulae as they stand, i.e., eluding
cancellation of the $k^2$ factors in Eq.~(\ref{int1}).
The $k$-integrals are then evaluated using
\begin{eqnarray}&&
\int~d^{d}k \exp\{ k^2 (x + y + z) + 2 k^{\mu} (x p_{\mu} + z q_{\mu})\}
\nonumber\\&&
= i \left(\frac{\pi}{x + y + z}\right)^{\frac{d}{2}}
\exp\left\{\frac{p^2 x y - m^2 z^2}{x + y + z}\right\},
\nonumber
\end{eqnarray}
\begin{eqnarray}&&
\int~d^{d}k ~k_{\alpha}
\exp\{ k^2 (x + y + z) + 2 k^{\mu} (x p_{\mu} + z q_{\mu})\} =
\nonumber\\&&
= i \left(\frac{\pi}{x + y + z}\right)^{\frac{d}{2}}
\exp\left\{\frac{p^2 x y - m^2 z^2}{x + y + z}\right\}
\left[- \frac{x p_{\alpha} + z q_{\alpha}}{x + y + z}\right],
\nonumber
\end{eqnarray}
\noindent
etc. up to six $k$-factors in the integrand.

From now on, all formulae will be written out for the sum
$$I \equiv I_{3(a)}(p,q) + I_{3(b)}(p,q).$$

Changing the integration variables $(x,y,z)$ to $(t,u,v)$ via
$$x = \frac{t (1 + t + u) v^2}{m^2 (1 + \alpha t u)},
~~y = \frac{u (1 + t + u) v^2}{m^2 (1 + \alpha t u)},
~~z = \frac{ (1 + t + u) v^2}{m^2 (1 + \alpha t u)},
~~\alpha \equiv - \frac{p^2}{m^2},$$
integrating $v$ out, subtracting the ultraviolet divergence
$$I^{{\rm div}} = \frac{1}{32\pi^2 \varepsilon}
\left(\frac{\mu}{m}\right)^{\varepsilon} E^{\mu\nu}(p)\eta_{\mu\nu}
\xi^2 (p^2 - 2 m^2),$$ setting $\varepsilon = 0$,
and retaining only the leading at $p^2 \to 0$ terms,
we obtain
\begin{eqnarray}&&\label{int2}
(I - I^{{\rm div}})_{{\varepsilon} \to 0}
\nonumber\\&&
=  \frac{E^{\mu\nu}(p)\xi}{16\pi^2}
\int_{0}^{\infty}\int_{0}^{\infty} du dt \left\{
\frac{8 m^2 (\xi - 1)}{p^2 D N^3}
\left(q_{\mu} q_{\nu} - \frac{m^2}{p^2} p_{\mu} p_{\nu} \right)
\left(6 - \frac{9}{D} + \frac{4}{D^2} \right)
\right.
\nonumber\\&&
\left.
+ \frac{\eta_{\mu\nu} m^2}{D N}
\left( 1 + \frac{12 \xi m^2}{N^2 p^2}
- \frac{8 \xi m^2}{D N^2 p^2} - \frac{5}{D} + \frac{4}{D^2}
\right)
\right\},
\nonumber\\&&
D \equiv 1 + \alpha u t, ~~N \equiv 1 + u + t.
\end{eqnarray}

Now, using Eqs.\ (\ref{ints}) of the Appendix, it is straightforward
to show that the remaining $(u,t)$-integral is zero:
\begin{eqnarray}
(I - I^{{\rm div}})_{{\varepsilon} \to 0} = 0.
\nonumber
\end{eqnarray}
\noindent
The first and the second lines in Eq.~(\ref{int2}) cancel independently
of each other, and so do the terms with and without $\xi$-factor in the
second line.

Thus, the $\xi$-independence of the one-loop $\hbar^0$ contribution
to the form factor Fig.~\ref{fig2}(b) is proved. As was explained in
Sec.~\ref{extension}, this fact allows us to consider this contribution
as describing deviations of the space-time metric from classical solutions
of the Einstein equations in the case when the gravitational field is
produced by only one particle of mass $m$. Now we can use the results of
the work \cite{donoghue} to determine the actual value of these deviations.
Restoring the ordinary units, with the help of Eqs.\ (55),(65)
of Ref.~\cite{donoghue}, and Eq.~(\ref{pr})\footnote{Note the notation
differences between Ref.~\cite{donoghue} and the present work.}
we obtain\footnote{The metric correction (\ref{main1}) itself turns out to be
$\xi$-independent. Therefore, there is no need in further investigation
of the $\xi$-dependence of observables, see the footnote 4.}
\begin{eqnarray}&&\label{main1}
h^{{\rm loop}}_{\mu\nu}(p) = - \frac{\pi^2 G^2}{c^2 \sqrt{-p^2}}
\left(3 m^2 \eta_{\mu\nu} + q_{\mu} q_{\nu}
+ 7 m^2\frac{p_{\mu} p_{\nu}}{p^2}\right).
\end{eqnarray}
\noindent
In particular, for the static field of a massive particle at rest, the
quantum correction to the time component of the metric, in the coordinate
space,
\begin{eqnarray}&&
h^{{\rm loop}}_{00}(r) = - \frac{2 G^2 m^2}{c^2 r^2}.
\nonumber
\end{eqnarray}
\noindent
Together with Eq.~(\ref{sch2}), this finally gives the following expression
for the first post-Newtonian correction to the gravitational potential of a
body with mass $M,$ consisting of $N$ identical (scalar) particles with
mass $m = M/N$
\begin{eqnarray}&&\label{main}
\Phi(r) =  - \frac{G M}{r} + \frac{G^2 M^2}{2 c^2 r^2}
- \frac{G^2 M^2}{N c^2 r^2}.
\end{eqnarray}

It remains only to make the following important remark. As was already noted
in Sec.~\ref{definition}, any attempt to give a self-contained notion of
gravitational (or any other) field has to start, despite its ultimate
purpose, with examination of matter interactions. Having restricted to zeroth
order in the Planck constant $\hbar,$ however, we made the explicit
introduction of additional matter playing the role of a measuring device
superfluous. Indeed, whatever kind of matter is considered, its quantum
corrected equations of motion to the order $\hbar^0$ are obtained simply
substituting the value (\ref{main}) in place of the potential entering
the ordinary classical equations.\footnote{Or, more generally, the value
$g^{clas}_{\mu\nu}+h^{loop}_{\mu\nu}$ in place of the classical metric,
where $g^{clas}_{\mu\nu}$ is the solution of the Einstein equations, and
$h^{loop}_{\mu\nu}$ is given by Eq.~(\ref{main1}).}
Any other quantum corrections due to nonlinearity of interaction of the
measuring apparatus with the gravitational field will be of higher orders
in the Planck constant. In other words, the value of the effective
gravitational field detected in the course of observation
of the apparatus motion will be just (\ref{main}).\footnote{There will be
$\hbar^0$ corrections due to quantum propagation of the measuring apparatus
also. We can get rid of them assuming small mass of the apparatus, just
like in the classical theory, see the footnote~1.}
This is why interpretation given in Sec.~\ref{extension} was formulated in
terms of the effective equations of motion of the gravitational, rather
than matter, field.

\section{Discussion and Conclusions}\label{conclude}

The gauge-independence of the gravitational form factors, underlying
our interpretation of the $\hbar^0$ loop contribution, is proved only
for a particular, though most important practically, choice of gauge.
Furthermore, the case of many-loop diagrams giving rise to the higher
post-Newtonian corrections has not been touched upon at all. On the other
hand, it is hardly believed that the gauge dependence cancellation
found out in Sec.~\ref{zero} is accidental. It takes place for any $\xi$
as well as for every independent tensor structure of diagram
Fig.~\ref{fig1}(b).

A very specific feature of this cancellation must be emphasized:
it holds only for the $\hbar^0$ part of the form factors. For instance,
the logarithmical part of $I$ (which is of the order $\hbar^1$) is not zero.
This raises the question of reasons the gauge dependence cancellation
originates from. The point is that the Slavnov identities, being valid
only for the totals of diagrams, cannot provide such reason, although
they do allow one to simplify calculation of the gauge-dependent parts of
diagrams.

As was shown in Sec.~\ref{extension}, the use of the effective action
is essential in establishing the correct correspondence between the
classical and quantum theories of gravity. This makes the problem of
physical interpretation of the effective action beyond the $\hbar^0$ order
even more persistent. It was mentioned in Sec.~\ref{definition} that
such interpretation can probably be given via the introduction of
point-particles into the theory. It should be noted, however, that
this approach cannot be justified from the first principles of quantum
theory, and therefore is out of our present concern. Nevertheless, one may
hope that the phenomenological approach advanced in Ref.~\cite{dalvit}
can be developed in a consistent manner, e.g., along the lines of
Refs.\ \cite{kazakov1,kazakov2}.

Finally, let us consider possible astrophysical applications of our
results. As we saw in Sec.~\ref{extension}, the loop contributions
to the post-Newtonian corrections are normally highly suppressed:
their relative value for the stars is of the order $10^{-56}-10^{-58}.$
Things are different, however, if an object consisting of strongly
interacting particles is considered. In this case, additivity of
individual contributions, implied in the course of derivation of
Eq.~(\ref{main}), does not take place. In the limit of infinitely-strong
interaction, the object is to be considered as a {\it particle}, i.e.,
one has to set $N = 1$ in Eq.~(\ref{main}). The post-Newtonian corrections
are then essentially different from those given by the classical
General Relativity.

An example of objects of this type is probably supplied by the black holes.
It should be noted, however, that the very existence of the horizon is
now under question. The potential $\Phi(r)$ may well turn out to be a
regular function of $r$ when all the $\hbar^0$ loop corrections are
taken into account.

\vspace{1cm}
\noindent
{\Large \bf Appendix}
\vspace{1cm}

The integrals
\begin{eqnarray}&&
J_{nm} \equiv \int_{0}^{\infty}\int_{0}^{\infty}
\frac{du dt}{(A + t + u)^n (B + \alpha t u)^m},
\nonumber
\end{eqnarray}
encountered in Sec.~\ref{zero}, can be evaluated as follows.
Consider the auxiliary quantity
\begin{eqnarray}&&
J(A,B) = \int_{0}^{\infty}\int_{0}^{\infty} \frac{du dt}{(A + t + u) (B + \alpha t u)},
\nonumber
\end{eqnarray}
\noindent
where $A,B>0$ are some numbers eventually set equal to 1.
Performing an elementary integration over $u,$ we get
\begin{eqnarray}&&
J(A,B) = \int_{0}^{\infty} dt
~\frac{\ln B - \ln \{\alpha t (A + t)\}}{B - \alpha t (A + t)}.
\nonumber
\end{eqnarray}

\begin{figure}
\hspace*{3cm}
\epsfxsize=10cm\epsfbox{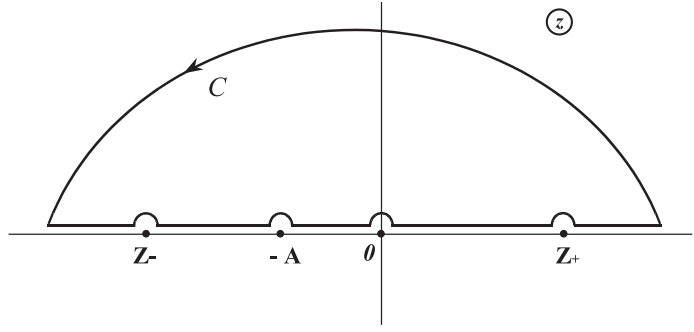}
\vspace{1cm}
\caption{Contour of integration in Eq.~(34)}
\label{fig3}
\end{figure}

Now consider the integral
\begin{eqnarray}&&
\tilde{J}(A,B) = \int_{C} dz f(z,A,B),
~~f(z,A,B) = \frac{\ln B - \ln \{\alpha z (A + z)\}}{B - \alpha z (A + z)},
\end{eqnarray}
taken over the contour $C$ shown in Fig.~\ref{fig3}. $\tilde{J}(A,B)$
is zero identically. On the other hand,
\begin{eqnarray}&&
\tilde{J}(A,B)
\nonumber\\&&
= \int^{-A}_{- \infty} dw
~\frac{\ln B - \ln \{\alpha w (A + w)\}}{B - \alpha w (A + w)}
+ \int_{-A}^{0} dw
~\frac{\ln B - \ln \{ - \alpha w (A + w)\} + i\pi}{B - \alpha w (A + w)}
\nonumber\\&&
+ \int_{0}^{+ \infty} dw
~\frac{\ln B - \ln \{\alpha w (A + w)\} + 2 i\pi}{B - \alpha w (A + w)}
- i\pi \sum\limits_{z_{+},z_{-}} {\rm Res} f(z,A,B),
\nonumber
\end{eqnarray}
\noindent
$z_{\pm}$ denoting the poles of the function $f(z,A,B),$
$$z_{\pm} = - \frac{A}{2} \pm \sqrt{\frac{B}{\alpha} + \frac{A^2}{4}}.$$
Change $w \to - A - w$ in the first integral.
A simple calculation then gives
\begin{eqnarray}&&\label{int4}
J(A,B)
= \frac{\pi^2}{2\sqrt{\alpha}} B^{-1/2}
\left(1 + \frac{\alpha A^2}{4 B}\right)^{-1/2}
- \frac{1}{2}\int_{0}^{A} dt
~\frac{\ln B - \ln \{\alpha t (A - t)\}}{B + \alpha t (A - t)}.
\end{eqnarray}

The roots are contained entirely in the first term on the right of
Eq.~(\ref{int4}). The integrals $J_{nm}$ are found by repeated
differentiation of Eq.~(\ref{int4}) with respect to $A,B$. Expanding
$(1 + \alpha A^2/4 B)^{-1/2}$ in powers of $\alpha,$
we find the leading terms needed in Sec.~\ref{zero}
\begin{eqnarray}&&\label{ints}
J^{{\rm root}}_{11} = \frac{\pi^2}{2\sqrt{\alpha}},
~~J^{{\rm root}}_{12} = \frac{\pi^2}{4\sqrt{\alpha}},
~~J^{{\rm root}}_{13} = \frac{3\pi^2}{16\sqrt{\alpha}},
\nonumber\\&&
J^{{\rm root}}_{31} = - \frac{\pi^2}{16}\sqrt{\alpha},
~~J^{{\rm root}}_{32} = - \frac{3\pi^2}{32}\sqrt{\alpha},
~~J^{{\rm root}}_{33} = - \frac{15\pi^2}{128}\sqrt{\alpha}.
\end{eqnarray}

\end{document}